\documentclass{ws-procs975x65}

\newcommand{\bra}[1]{\left\langle #1 \right|}
\newcommand{\ket}[1]{\left| #1 \right\rangle}
\def\figsubcap#1{\par\noindent\centering\footnotesize(#1)}

\begin{document}

\title{The scalar spectrum of many-flavour QCD\footnote{Talk given at
    KMI-GCOE Workshop on Strong Coupling Gauge Theories in the LHC
    Perspective (SCGT 12), KMI, Nagoya University, December 4-7, 2012.}}

\author{Yasumichi~Aoki$^a$, Tatsumi~Aoyama$^a$, Masafumi~Kurachi$^a$,
   Toshihide~Maskawa$^a$, Kei-ichi~Nagai$^a$, Hiroshi~Ohki$^a$,
   Enrico~Rinaldi$^{a,b}$\footnote{Speaker. E-mail: e.rinaldi@sms.ed.ac.uk},
   Akihiro~Shibata$^c$, Koichi~Yamawaki$^a$, Takeshi~Yamazaki$^a$\\ \vspace{3pt} LatKMI collaboration}
\address{$^a$Kobayashi-Maskawa Institute for the Origin of Particles and the
  Universe (KMI),\\ Nagoya University, Nagoya, 464-8602, Japan\\
  $^b$SUPA and School of Physics and Astronomy, University of
  Edinburgh,\\Edinburgh, EH9 3JZ, UK\\
  $^c$Computing Research Center, High Energy Accelerator Research
  Organization (KEK),\\ Tsukuba 305-0801, Japan}

\begin{abstract}
The LatKMI collaboration is studying systematically the dynamical properties of
$N_f = 4,8,12,16$ SU(3) gauge theories using lattice simulations with
(HISQ) staggered fermions. Exploring the spectrum of many-flavour QCD,
and its scaling near the chiral limit, is mandatory in order to
establish if one of these models realises the Walking Technicolor
scenario. Although lattice technologies to study the mesonic
spectrum are well developed, scalar flavour-singlet states still
require extra effort to be determined. In addition, gluonic
observables usually require large-statistic simulations and powerful
noise-reduction techniques. In the following, we present useful
spectroscopic methods to investigate scalar glueballs and scalar
flavour-singlet mesons, together with the current status of the scalar
spectrum in $N_f = 12$ QCD from the LatKMI collaboration.
\end{abstract}

\keywords{Lattice Gauge Theories, Walking Technicolour, Glueballs}

\bodymatter
\section{Introduction}
\label{sec:intro}

It has been known for more than 30 years that the electroweak symmetry
breaking (EWSB) mechanism could be due to a new strongly-interacting
sector at energies of $\mathcal{O}$(TeV). Since then, this idea has
evolved to become one of the most promising scenarios for physics
beyond the Standard Model, the so-called Walking Technicolour with a
large anomalous dimension $\gamma_m \sim 1$ and an approximate scale
invariance~\cite{Yamawaki:1985zg} (See also similar
works~\cite{Holdom:1984sk,Appelquist:1986an,Akiba:1985rr} without
notion of  anomalous dimension and scale invariance).\\
Recently, the discovery of a new particle of mass $\sim125$ GeV at
the Large Hadron Collider~\cite{:2012gu,:2012gk} has attracted a lot
of attention and it is
now a pressing issue to determine its properties and its nature. The
new particle's branching ratios are consistent with the ones of the
Standard Model Higgs boson, but the very interesting
possibility of this particle being of composite nature, a
techni-dilaton~\cite{Yamawaki:1985zg}, as predicted by Walking
Technicolour models is also
viable~\cite{Matsuzaki:2012vc,Matsuzaki:2012xx}. The LHC experiments
will
be working very hard in the following years, trying to assess the
validity of their analysis for the different decay channels of the new
particle. At the same time, the effort of the theory community will go
in a similar direction; new increasingly precise lattice simulations
might help addressing the question whether or not a more fundamental
strongly-interacting gauge sector is responsible for the origin of
mass (a review of recent lattice studies can be found in
Refs.~\refcite{Neil:2012cb,Giedt:2012hg}).\\
The LatKMI collaboration is focusing on a systematic study of SU(3)
gauge theories with different numbers $N_f$ of fundamental massless
fermions using lattice Monte Carlo simulations (see
Refs.~\refcite{Aoki:2012eq,Aoki:2013dz,Aoki:2012yd} for recent updates). The main
purpose of such on-going project is to find a near-conformal theory with
a large anomalous dimension $\gamma_m \sim 1$ by investigating the
infrared dynamics of the theory which is in the non-perturbative
domain. If such a theory will be found, the important question will
be to identify the composite scalar state, techni-dilaton, mimicking the role of the
\emph{Higgs} boson of the Standard Model and to study its mass as it
arises from a purely non-perturbative dynamics.\\
The study of scalar single-particle resonances on the lattice is
complicated due to the poor quality of the signal-to-noise
ratio in Monte Carlo simulation. Even in state-of-the-art lattice QCD
calculations, 
where an extreme degree of accuracy has been reached in recent years
for many physical observables, the spectrum of scalar mesons is still
an open issue. Moreover, scalar excitations
of gluonic degrees of 
freedom (glueballs) introduce an additional factor of complexity since
they can mix with fermionic states: a careful study is necessary to
distinguish the nature of states in the scalar channel (for an up to
date review see Ref.~\refcite{Ochs:2013fk}). In the
following, we describe the status of LatKMI investigation of the
scalar spectrum from lattice simulations of $N_f=12$ QCD. The data
shown in the following are preliminary.

\section{Lattice simulations}

\begin{figure}
  \begin{center}
    \psfig{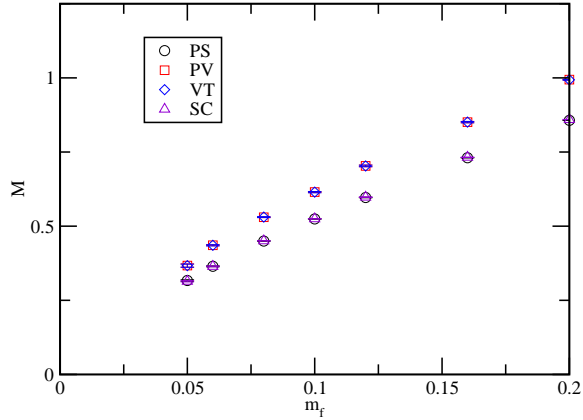}
    \caption{An example of the small taste splitting in the spectrum of
      $N_f=12$ QCD with HISQ fermions at
      $\beta=4.0$. The figure is taken from
      Ref.~\refcite{Aoki:2012eq}. M and $m_f$ are the 
      meson mass and the bare quark mass respectively, both in units of
      the common fixed lattice spacing $a$. PS and SC refer to
      pseudoscalar mesons coming
      from operators with different internal taste structure, whereas
      PV and VT refer to vector mesons.}
    \label{fig:tastesplit}
  \end{center}
\end{figure}

The lattice gauge configurations for the SU(3) theory with $N_f=12$
fundamental flavours are generated by the HMC algorithm
using a tree-level Symanzik gauge action and the HISQ (Highly Improved
Staggered Quark~\cite{Follana:2006rc}) action
for the dynamical fermions. The flavour symmetry breaking of this
action (measured as taste splitting) is highly suppressed in
QCD~\cite{Bazavov:2011nk} and 
it is almost negligible for the theory we investigate. At the
couplings used in this study, the masses 
of bound states in the same taste multiplet can not be distinguished
from one another as shown in Fig.~\ref{fig:tastesplit}. More
details on the calculated mesonic spectrum and decay constants can be
found in Ref.~\refcite{Aoki:2012eq}.

At fixed lattice spacing defined by the bare coupling constant
$\beta=6/g^2=4.0$, we simulate two physical volumes $V=(aL)^3$ with
$L=18, \, 24$ and aspect ratio $T/L = 4/3$. On the smaller volume we investigate the scalar
spectrum at five different bare quark masses $0.06 < am_f < 0.16 $;
the bare mass changes by almost a factor of 3 and this allows us
to identify a possible quark mass dependence in the measured
observables. Another important feature of our simulation is the large
number of Monte Carlo trajectories obtained from uninterrupted Markov
chains after thermalization. For any set of parameters
explored, we collect more than 15000 trajectories (with MD step set to unity),
which is necessary to contrast the rapid degradation of signal to
noise ratio of the scalar correlators. A summary
of the simulated parameters is found in Tab.~\ref{tab:simulations}.

\begin{table}
\tbl{Parameters of lattice simulations for $N_f=12$ QCD at fixed
  $\beta=4.0$. $N_{\rm traj}$ is the number of Monte Carlo trajectories after
  thermalisation, $N_{\rm cfgs}$ is the number of saved gauge
  configurations. For different ensembles, configurations are saved
  every 2 or 5 trajectories.}
{\begin{tabular}{p{2cm}p{2cm}p{2cm}p{2cm}}
\toprule
$L \times T$ & $am_f$ & $N_{\rm traj}$ & $N_{\rm cfgs}$  \\\colrule
$18 \times 24$ & 0.06 & 15445 & 3090 \\
$18 \times 24$ & 0.08 & 22200 & 4440 \\
$18 \times 24$ & 0.10 & 15300 & 3060 \\
$18 \times 24$ & 0.12 & 21500 & 4300 \\
$18 \times 24$ & 0.16 & 18450 & 3690 \\
$24 \times 32$ & 0.06 & 28000 & 14000 \\
$24 \times 32$ & 0.08 & 19280 & 9640 \\
$24 \times 32$ & 0.10 & 18960 & 9480 \\\botrule
\end{tabular}}
\label{tab:simulations}
\end{table}

\section{Glueball measurement}

\begin{figure}[t]
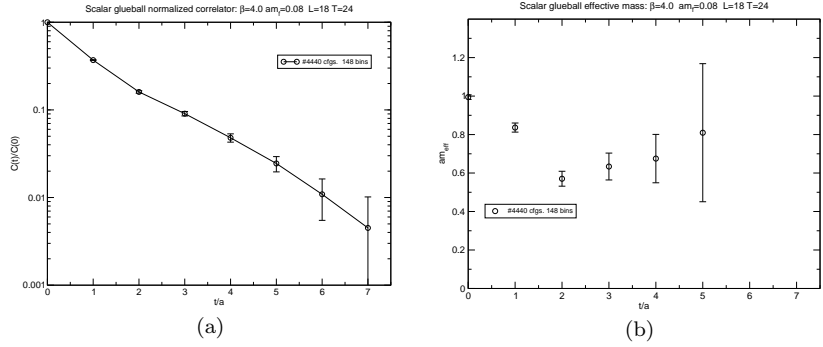
%
  \begin{center}
    \parbox{2.1in}{\epsfig{figure=corrl18m08.eps,width=2in}
      \figsubcap{a}}
    \hspace*{4pt}
    \parbox{2.1in}{\epsfig{figure=meffl18m08.eps,width=2in}
      \figsubcap{b}}
    \caption{Analysis of $4440$ configurations for $\beta=4.0$ and
      $am_f=0.08$ on the smaller volume.
      (a) The time dependence of the normalized scalar correlator in
      logarithmic scale; the linear behaviour corresponds to the
      propagation of a single state (b) The effective mass of the
      corresponding correlator; the signal is lost after 6 timeslices,
      but the plateaux is already present at short temporal distances.} 
    \label{fig:l18m08}
  \end{center}
\end{figure}

\begin{figure}[t]
  \begin{center}
    \psfig{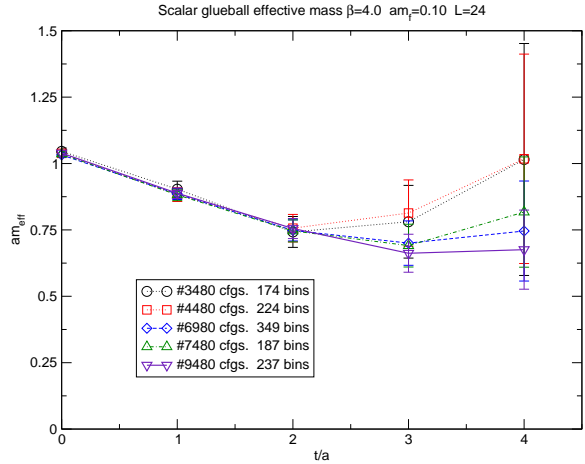}
    \caption{An example of the scalar effective mass at $\beta=4.0$
      and $am_f=0.10$ on the larger volume $L\times T=24 \times 32$ as
      the number of configuration is increased. A stable plateaux
      becomes more and more visible as the statistics increases.}
    \label{fig:statistics}
  \end{center}
\end{figure}

It has been pointed out~\cite{DelDebbio:2009fd} that signs of
infrared conformality can be found by comparing the mesonic spectrum
with the gluonic one. Such a comparison was done in
the SU(2) gauge theory with two flavours of adjoint
fermions~\cite{DelDebbio:2010hx} and the results supported the
hypothesis of an infrared fixed point being responsible for the
dynamics of the theory on the critical massless surface.\\
Addressing a measurement of the glueball spectrum is intrinsically
difficult due to the poor signal over noise ratio in gluonic
correlators. 
The theory behind the
measurement of glueball masses in lattice spectroscopy has been known
for a long time~\cite{Berg:1982kp}. Recently, a powerful framework to
measure gluonic bound states has been 
developed~\cite{Lucini:2010nv} that improves upon previous results.
The method employs a large number of different operators
built from gauge-invariant combinations of gauge links in such a way
that a robust basis for a variational ansatz can be created.
The variational approach, with such a large basis, has been shown to
produce excellent results for different gauge
theories~\cite{Lucini:2010nv,DelDebbio:2012mr,Gregory:2012hu}. In the
following we briefly
review the technology involved in such a variational method and we
show the preliminary results obtained by the LatKMI collaboration
using the gauge configurations in Tab.~\ref{tab:simulations}. This is
the first time that
glueball states are investigated in many-flavour SU($3$) gauge theory.\\
Eigenstates of the lattice Hamiltonian fall into irreducible
representations of the symmetry group characterising the four-dimensional lattice
system. In particular, there are $20$ symmetry channels for glueballs
on a three-dimensional spatial lattice, corresponding to the $5$
irreducible representations of the cubic group and including both
parity and charge quantum numbers. In the following, we focus on the
scalar $0^{++}$ channel that is expected to feature the lightest state. 
On each spatial point of the
lattice we build gauge-invariant interpolating operators
$\mathcal{O}_{\alpha}(x,t)$ with well-defined 
rotational quantum number by using a prescribed linear combination of
traced spatial Wilson loops. Zero-momentum operators consist in an average
over the whole volume $V$. By using differently shaped Wilson loops
we construct $32$ different basis operators for
the scalar glueball
(each of these operators come in different blocking sizes and they are
all included in the variational basis).\\
Let us now give an example of a typical glueball measurement. For every
configuration we measure a matrix of correlation functions, whose
elements are
\begin{equation}
  \label{eq:corr-matrix}
  \tilde{C}_{\alpha\beta}(t) \; = \; \sum_\tau \bra{0}
  \mathcal{O}_{\alpha}^{\dag} (t+\tau) \mathcal{O}_{\beta}(\tau) \ket{0}
  \, .
\end{equation}
By solving the generalised eigenvalue problem for the matrix above,
optimal operators (i.e. those that create almost pure states $\ket{i}$)
can be found that are a linear sum of the basis vectors
\begin{equation}
  \label{eq:optimal-op}
  \tilde{\mathcal{O}}_{i} (t) \; = \; \sum_\alpha v^i_\alpha
  \mathcal{O}_\alpha(t)
  \, ; \qquad
  \tilde{\mathcal{O}}_{i} (t)\ket{0} \approx \ket{i}
  \, ,
\end{equation}
where $v^i_\alpha$ are the components of the $i^{\rm th}$
eigenvector. Different eigenvectors $v^i$ correspond to different states.
The mass $m_i$ of the $i^{\rm th}$ state is then extracted by fitting
correlators of optimal operators using
\begin{equation}
\label{eq:coshFIT}
\bar{C}_{ii}(t) = |c_i|^2 \left( e^{-m_{i}t} + e^{-m_{i}(T-t)}\right)
\, ,
\end{equation}
where $T$ is the length of the lattice in the time
direction and the functional form is a consequence of the
usual exponential decay in a lattice with periodic boundary condition
in the time direction.\\
In general, glueball correlators are very noisy and this limits the
usefulness of numerical correlators to short time separations. However,
although Eq.~(\ref{eq:coshFIT}) is only valid at large $t$, if the
overlap with an Hamiltonian state is almost perfect, it is possible to
extract a reliable value for the mass at short time separation, since
the decay is largely dominated by a single state. For this to be true, a
careful construction of the variational basis is paramount. Whether
the state created by an optimal operator $\tilde{\mathcal{O}}_{i} (t)$ is
a good approximation of the Hamiltonian 
eigenstates can be checked by looking at the value of the overlap $|c_i|^2$:  the
closer this number is to one, the better is the variational calculation.\\
We measure the correlators of scalar gluonic operators on the
configurations of Tab.~\ref{tab:simulations}. The measurements are
taken for every configurations and statistical errors are estimated
using jack-knife. We check for autocorrelations by changing the size
of the jack-knife bins.\\
For each set of simulated parameters, we perform the variational
analysis described above and we obtain the correlator corresponding to
the optimal operator that projects on the ground state
($i=1$). In some cases, the correlator for the first excited
state ($i=2$) is studied as well in order to estimate the higher
states contamination in the ground state estimate. We calculate the
effective mass and we 
look for a plateaux where Eq.~(\ref{eq:coshFIT}) can be applied. An
example of the scalar ground state correlator and its effective mass
is shown in Fig.~\ref{fig:l18m08}.\\
The high number of analysed configurations is very helpful when
determining the effective mass plateaux. For example
Fig.~\ref{fig:statistics} shows how 
the effective mass changes when the statistic is increased.
\section{Flavour-singlet scalar meson}

\begin{figure}[t]
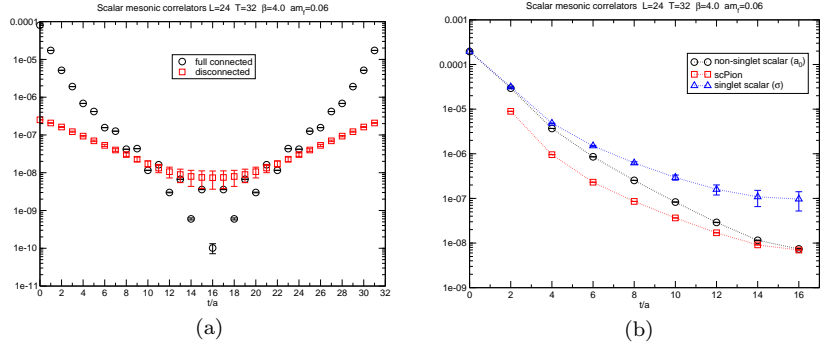
%
  \begin{center}
    \parbox{2.1in}{\epsfig{figure=corrb4l24m06-mes.eps,width=2in}
      \figsubcap{a}}
    \hspace*{4pt}
    \parbox{2.1in}{\epsfig{figure=corrb4l24m06-mes-par.eps,width=2in}
      \figsubcap{b}}
    \caption{Results for the analysis of scalar mesonic interpolating operators
      (a) The full connected and disconnected correlators where
      the vev as been subtracted in the latter. The connected
      correlator still includes the parity partner contributions. (b) The correlators of
      the different channels extracted from $C(t)$, where disconnected
      terms have been included for the singlet case only.}
    \label{fig:mes-results}
  \end{center}
\end{figure}

\begin{figure}[b]
  \begin{center}
    \psfig{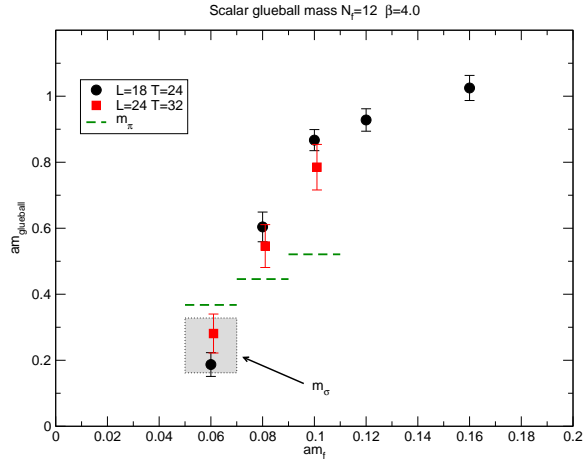}
    \caption{Summary of the fitted scalar glueball masses for $N_f=12$
      QCD at $\beta=4.0$ and for several bare fermion masses. Some
      values of the pion mass are shown for comparison. The point on
      different volumes have been slightly displaced for clarity. The
      grey box indicates the location of the scalar flavour-singlet
      mass $m_{\sigma}$ on the ensemble with the lightest bare quark mass.}
    \label{fig:summary}
  \end{center}
\end{figure}

In the staggered discretization, fermionic fields defined on the
lattice sites have an additional {\it taste} structure. The scalar $0^+$
fermionic interpolating operator that we used in the construction of
our propagators is the simplest one and has a
$(1\times1)$ spin--taste content. As is generally the case, this
operator couples also to a negative parity $0^-$ state whose contribution
must be accounted for in the analysis of the full connected correlator
$C(t)$. However,
for the scalar flavour-singlet case (referred to as $\sigma$ in the
following), we need to compute
disconnected diagrams and the vacuum expectation value of the mesonic
operator.\\
Measuring disconnected contributions to mesonic observables is very
challenging and ad-hoc noise reduction techniques have been developed
and used successfully in lattice QCD (LQCD). We used the stochastic source
method~\cite{Venkataraman:1997xi,Gregory:2007ev} with Gaussian noise
and local source and sink insertions. The number of noise vectors for
each configuration has been chosen to the minimal value that yields
the asymptotic value of the disconnected correlator $D(t)$ on the available
statistics.\\
Thanks to the relative high number of configurations analysed at the
lightest bare fermion mass (cfr. Tab.~\ref{tab:simulations}), we are
able to extract a clean signal for the disconnected correlator $D(t)$
at large temporal distances and including the vacuum expectation value
subtraction, as it is shown in
Fig.~\ref{fig:mes-results}(a). The scalar singlet correlator $C_{\sigma}(t)$ is
extracted from combinations of $C(t)$ and $D(t)$ that
properly account for the
effects of the parity partner meson and of the vacuum fluctuations:
\begin{align}
  \label{eq:mes-corr}
  \tilde{C}(2t) & \; = \; 2C(2t) + C(2t+1) + C(2t-1) \, ,\\
  \tilde{D}(2t) & \; = \; 2D(2t) + D(2t+1) + D(2t-1) \, ,\\
  C_{\sigma}(2t) & \; = \; - \tilde{C}(2t) + 3 \tilde{D}(2t) \, .
\end{align}
The factor of $3$ in the last equation comes from the number of actual
fermions in the simulation ($3$ staggered fermions $\times$ $4$ tastes
= $12$ flavours). From the full connected correlator, the
non-singlet ($a_0$) and the pseudoscalar (scPion) states can be
isolated as well. The 
three different channels are compared in Fig.~\ref{fig:mes-results}(b)
and the results strongly indicates that the $\sigma$ meson mass is
smaller than the pseudoscalar one. Moreover, although the
scalar effective mass plateaux appears only at very large temporal
separation and the statistical error are large, the corresponding mass
is compatible with the one obtained using only purely gluonic
operators. The comparison between the gluonic scalar spectrum and the
fermionic scalar singlet is reported in Fig.~\ref{fig:summary}.
\section{Conclusions}

We performed a study of the scalar glueball and
scalar flavour-singlet fermionic state. Advanced lattice
spectroscopy techniques that have proven successful in LQCD studies
have been applied for the first time to explore the scalar spectrum of
many-flavour QCD.\\
The results shown in the previous sections clearly show that the
variational method used to extract masses of gluonic bound states is
very efficient and allowed us to study the qualitative quark mass
dependence of scalar glueballs. Although the results are preliminary
and the statistical errors of the order $\sim 10\%$, it is very
encouraging that a signal for an effective mass plateaux could be
obtained even for relatively small time separation. As for the scalar
fermionic correlator including the disconnected term,
although an effective mass plateaux is observed at somewhat larger time
due to the lack of improved operators, it is remarkable that the effective
mass is consistent with the mass obtained from the gluonic operators.\\
The most striking feature of the measured scalar spectrum is the
appearance of a state lighter than the pion for low bare quark
masses as it is shown in Fig.~\ref{fig:summary}. Such a state appears
both in gluonic and fermionic
correlators, indicating that mixing between gluons and fermions plays
a relevant role. A future
variational analysis including both kind of operators described in the
previous sections will help disentangle mixed contributions.
Moreover, a careful finite size study is under completion and possible
systematic effects of the lattice regularization are being investigated.\\
Despite it being studied by several groups using different approaches, $N_f=12$
QCD has not yet been identified as an infrared-conformal or near-conformal
theory. The majority of studies suggest the presence of an infrared
fixed point with a small anomalous dimension. If this turns out to
be the case, such a theory would not be a viable candidate for a
phenomenologically interesting Walking Technicolour model. A walking
theory, which may be realised near the edge of the conformal window,
should have its property influenced by the near-conformal
theory. The light scalar state observed for $N_f=12$ in this study is a
promising signal for the search of a successful  walking theory, where
the existence of a light scalar is now required by LHC experiment. 
A possible future direction is to look at more viable candidates for 
Walking Technicolour  models; for example, it will be interesting
to investigate the scalar spectrum of the $N_f=8$ SU(3) theory
where, if the theory has a walking regime, the scalar state could be
identified with the pseudo-NG boson coming from the breaking of
conformal symmetry.

\section*{Acknowledgments}
Numerical simulation has been carried out on the supercomputer 
system $\varphi$ at KMI in Nagoya university,
and on the computer facilities of the Research Institute for Information
Technology at Kyushu University.\\
This work is supported by the JSPS Grant-in-Aid for 
Scientific Research (S) No.22224003, (C) No.23540300 (K.Y.) 
and (C) No.21540289 (Y.A.),
and also by Grants-in-Aid of the Japanese Ministry for Scientific Research 
on Innovative Areas No. 23105708 (T.Y.). E.R. is currently supported
by a SUPA Prize Studentship and was supported by a
FY2012 JSPS Postdoctoral Fellowship for Foreign Researchers
(short-term), ID PE12518.

\bibliographystyle{ws-procs975x65}
\bibliography{technicolor_biblio}

\end{document}